\documentclass{optica-article}

\journal{opticajournal} 

\articletype{Research Article}

\usepackage{pifont}
\usepackage{array,tabularx,booktabs}
\usepackage{multirow}
\usepackage{stfloats}
\usepackage{booktabs}
\usepackage{makecell}
\newcolumntype{Y}{>{\centering\arraybackslash}X} %
\usepackage{lineno} 
\newcommand{\rev}[1]{\textcolor{black}{#1}}
\newcommand{\rrev}[1]{\textcolor{black}{#1}}

\begin{document}

\title{Link-adaptive digital twin for robust physical-layer modeling in hybrid-amplified ultra-wideband optical networks}

\author{Xiaoxuan Gao,\authormark{1} Rentao Gu,\authormark{1,*} Yingchun Wang,\authormark{2} Xinyi Liu,\authormark{1} Junshi Gao,\authormark{2} and Yuefeng Ji\authormark{1}}

\address{\authormark{1}State Key Lab of Information Photonics and Optical Communications, Beijing University of Posts and Telecommunications, Beijing, 100876, China\\
\authormark{2}China Mobile Group Design Institute Co., Ltd., Beijing, 100080, China\\}

\email{\authormark{*}rentaogu@bupt.edu.cn} 


\begin{abstract*} 
Ultra-wideband optical networks represent a practical solution for expanding communication capacity and supporting emerging applications such as artificial intelligence data center interconnection and 6G-oriented intelligent networks. Accurate physical-layer modeling has become increasingly essential to ensure reliable ultra-wideband network operation and capacity optimization, particularly under the intensified inter-channel stimulated Raman scattering (ISRS) effect. This paper proposes the link-adaptive digital twin (LA-DT) of hybrid-amplified ultra-wideband links to overcome the generalization and speed limitations of existing modeling methods, achieving accurate physical-layer modeling as well as robust generalized signal-to-noise ratio (GSNR) estimation across diverse links. First, to address the heterogeneity of EDFAs, the GSNR modeling task is decomposed into three key power predictions, including amplified spontaneous emission (ASE), nonlinear interference (NLI), and signal powers before entering the EDFA. Second, to enhance generalization in cross-scenario power prediction, three dedicated DT models are developed based on a novel neural network architecture that introduces linear modulation layers (LMLs). Third, to enable rapid adaptation to previously unseen scenarios with only limited data, three domain discriminators are designed to guide the few-shot fine-tuning of the LMLs. \rev{Finally, the proposed LA-DT explicitly accounts for insertion loss induced by Raman amplifiers (RAs), thereby improving modeling reliability under practical deployment conditions.} Results demonstrate that LA-DT achieves accurate physical-layer modeling across 35 scenarios with diverse fiber lengths, launch powers, Raman pump powers, and insertion losses. Compared with the baseline method, LA-DT reduces the average root mean square error (RMSE) of NLI, ASE, and signal power predictions to 0.151 dBm, 0.111 dB, and 0.113 dBm, respectively, corresponding to improvements of 56.0\%, 58.4\%, and 52.7\%, and achieves an average GSNR estimation RMSE as low as 0.114 dB, representing a 55.8\% improvement. Moreover, for 12 previously unseen scenarios, LA-DT maintains high accuracy through domain discriminator–guided few-shot fine-tuning with only 20 samples per scenario, achieving an average GSNR estimation RMSE of 0.159 dB and demonstrating strong adaptability and robustness.

\end{abstract*}

\section{Introduction}
\label{Sec1}
The rapid development of emerging technologies such as intelligent computing, 6G communications, and massive machine-type communications has increased the demand for high-capacity optical networks, creating an urgent need for capacity expansion \cite{1,13}. 
Ultra-wideband optical networks and space-division multiplexing (SDM) have received considerable attention as two mainstream solutions for capacity expansion. Compared with SDM, ultra-wideband expansion provides a more cost-effective and scalable approach by exploiting new bands beyond the conventional C-band without additional fiber deployment\cite{2,3}, making it a practical option for near-term network upgrades \cite{4,5,6,7,8}. 

\subsection{Motivation}
Accurate physical-layer modeling is critical for reliable network operation and capacity optimization \cite{16,17,18}, particularly in ultra-wideband optical networks due to the intensified ISRS effect \cite{19}. In ultra-wideband optical networks, physical-layer modeling must address several critical challenges. First, the ISRS effect becomes non-negligible with the introduction of new bands, leading to power transfer from high-frequency to low-frequency channels and degrading transmission performance. Second, hybrid-amplified links combining Raman amplifiers (RAs) and doped fiber amplifiers (DFAs) are widely deployed \cite{10,11} to mitigate lightpath degradation \cite{12} and enhance overall performance \cite{14,15}, which introduces complex pump–signal interactions that further complicate physical-layer modeling. Third, link parameters such as fiber length, launch power, Raman pump power, and insertion loss vary significantly across different transmission scenarios, requiring robust models that maintain high accuracy under diverse conditions. Finally, practical network operation requires real-time feedback, demanding modeling approaches that are both accurate and computationally efficient.

However, existing modeling approaches struggle to simultaneously meet the requirements of accuracy, computational efficiency, and cross-scenario generalization. Analytical models, such as GN-based extensions, provide valuable theoretical insights and are widely recognized for their accuracy. However, they often involve high computational complexity, especially as the number of channels increases with the introduction of new bands, and depend on precise link parameter measurements, which are often difficult to obtain in practice. Machine learning (ML)–based methods have emerged as a complementary direction, offering fast inference and the capability to capture complex nonlinear effects. However, conventional ML models often generalize poorly across diverse link configurations, as their performance is highly sensitive to variations in link parameters. There is a practical necessity for a generalized modeling approach that can achieve precise physical-layer modeling, support fast inference for real-time operation, and provide robust adaptation to diverse link conditions.

\subsection{Contribution}
In this paper, we propose the link-adaptive digital twin (LA-DT) that enables robust physical-layer modeling in hybrid-amplified ultra-wideband links, addressing key modeling challenges of limited generalization, high computational complexity, and slow inference speed. LA-DT is capable of adapting to diverse link conditions, accounting for variations in fiber length, launch power, Raman pump power, and insertion losses. In addition, it explicitly considers the variation of insertion loss induced by Raman pumps for the first time, a factor that is often overlooked but has a non-negligible impact on practical network performance. The proposed LA-DT incorporates the following key innovations:

\textbullet\textbf{ EDFA heterogeneity-driven decomposed GSNR modeling:} To mitigate the accuracy degradation of GSNR modeling caused by EDFA heterogeneity in practical ultra-wideband networks, we adopt a decomposed strategy that predicts three key physical-layer powers before entering EDFA, rather than modeling GSNR directly. This approach ensures accurate GSNR estimation under varying EDFA gain and noise figure (NF) settings, while enhancing interpretability across diverse link conditions.

\textbullet\textbf{ DeepModNet-based DT architecture:} To enhance generalization in cross-scenario power prediction, we developed three dedicated DT models, DeepModNet-NLI, DeepModNet-ASE, and DeepModNet-Sig. These DeepModNet models are built on a novel neural network architecture that introduces linear modulation layers (LMLs), enabling dynamic conditional modeling through a shared backbone and cross-domain regression. Accurate prediction of three key powers is achieved by the constructed DT models with strong generalization across diverse link conditions, including variations in Raman pump power, launch power, fiber length, Raman pump insertion loss, and fiber insertion loss.  Compared with the baseline method, DeepModNet-based models improve the prediction accuracy of NLI, ASE, and signal power by 56.0\%, 58.4\%, and 52.7\%, across 35 different scenarios, respectively, resulting in a 55.8\% improvement in GSNR estimation accuracy, with an average root mean square error (RMSE) as low as 0.114 dB .

\textbullet\textbf{ Domain discriminator–guided few-shot fine-tuning:} To further support rapid adaptation to previously unseen scenarios with only limited data, three domain discriminators are designed to guide the few-shot fine-tuning of the DeepModNet-based DT models. Results show that, fast adaptation to unseen link conditions is achieved under the guidance of the designed discriminators. Using only 20 new samples, the prediction accuracy of the NLI, ASE, and signal power improves by 67.7\%, 69.0\%, and 64.8\%, respectively, while the average GSNR estimation RMSE is reduced to 0.159 dB, representing a 61.2\% improvement.

The rest of the paper is organized as follows. We first summarize related works in Section~\ref{related works}. Then, Section~\ref{Sec3} formulates the physical-layer modeling of hybrid-amplified ultra-wideband optical networks and outlines the associated challenges. Section~\ref{Sec4} presents the proposed LA-DT, including the overall workflow, the architecture of the DeepModNet-based DT models and the domain discriminator–guided fine-tuning mechanism. Section~\ref{Sec5} presents a comprehensive analysis and discussion of the results, covering baseline performance comparison, accuracy of power prediction and GSNR estimation, and evaluations of generalization and adaptation performance across diverse scenarios. Finally, the conclusions of this work are presented in Section~\ref{Sec6}.

\section{Related works}
\label{related works}
Recently, extensive research has focused on various aspects of physical-layer modeling in ultra-wideband optical networks, including signal power evolution in presence of ISRS, ASE noise accumulation, NLI generation, RA gain prediction and optimization \cite{9}, EDFA characteristics, and GSNR estimation. The mainstream modeling approaches can be categorized into analytical models and machine learning–based models. Analytical models, such as extensions of the Gaussian Noise (GN) model, provide theoretical insights with high accuracy but are often limited by their computational complexity and dependence on precise parameter measurement. Machine learning–based models, offer fast online inference and strong nonlinear fitting capabilities, while their generalization across diverse link conditions remains a major challenge. Digital twin (DT) technology has recently been introduced in optical networks \cite{17, 10946004}, offering a unified framework to construct high-fidelity digital replicas of physical-layer behaviors. By enabling accurate modeling of link characteristics, DT models provide the foundation for reliable performance prediction and further support intelligent network optimization, with applications in quality of transmission (QoT) estimation, dynamic resource allocation, and fault diagnosis.

\begin{table*}[!t]
\centering
\captionsetup{width=\linewidth}
\caption{Summary of related physical-layer modeling methods in ultra-wideband optical networks}
\label{tab:relatedworks}
\renewcommand{\arraystretch}{1.1}
\setlength{\tabcolsep}{2.5pt}
\scriptsize

\begin{tabularx}{\linewidth}{
  >{\centering\arraybackslash}p{1cm}  
  >{\centering\arraybackslash}p{2.0cm}   
  >{\centering\arraybackslash}p{1.95cm}  
  >{\centering\arraybackslash}p{1.5cm}  
  >{\centering\arraybackslash}p{1cm}  
  *5{Y}                                   
}
\toprule
\multirow{2}{*}{\textbf{Paper}} &
\multirow{2}{*}{\textbf{Methods}} &
\multirow{2}{*}{\makecell{\textbf{Modeling}\\\textbf{Objective}}} &
\multirow{2}{*}{\makecell{\textbf{Accurate}\\\textbf{Parameter}\\\textbf{Measurement}}} &
\multirow{2}{*}{\makecell{\textbf{Real-time}\\\textbf{Inference}}} &
\multicolumn{5}{c}{\textbf{Generalization}} \\
\cmidrule(lr){6-10}
 & & & & &
 \makecell{\textbf{Launch}\\\textbf{Power}} &
 \makecell{\textbf{Fiber}\\\textbf{Length}} &
 \makecell{\textbf{Loss}} &
 \makecell{\textbf{Pump}\\\textbf{Power}} &
 \makecell{\textbf{EDFA}\\\textbf{Config.}} \\
\midrule
\cite{20}         & PINN model                                 & Power evolution prediction                              & Not Required & \ding{51} & \ding{51} & \ding{51} & \ding{55} & --        & \ding{55} \\
\cite{10647400}   & NN-based regressors                         & Power \& GSNR estimation                                & Not Required & \ding{51} & \ding{51} & \ding{51} & \ding{55} & --        & \ding{55} \\
\cite{21}         & ANN-based models                            & Power profile prediction                                & Not Required & \ding{51} & \ding{51} & \ding{51} & \ding{55} & \ding{55} & \ding{55} \\
\cite{23}         & Load-aware NN model                         & RA gain prediction                                      & Not Required & \ding{51} & \ding{51} & \ding{55} & \ding{55} & \ding{51} & --        \\
\cite{9239870}    & ML-based inverse mapping model              & RA gain design                                          & Not Required & \ding{51} & \ding{55} & \ding{55} & \ding{55} & \ding{51} & \ding{55} \\
\cite{24}         & Fiber-agnostic NN model                     & RA gain prediction                                      & Not Required & \ding{51} & \ding{55} & \ding{51} & \ding{55} & \ding{51} & --        \\
\cite{27}         & NN-based DT \& online optimization          & RA gain prediction \& optimization                       & Not Required & \ding{51} & \ding{55} & \ding{55} & \ding{55} & \ding{51} & \ding{55} \\
\cite{25}         & Input-parameter refinement-assisted DT model& RA modeling                                             & Not Required & \ding{51} & \ding{51} & \ding{55} & \ding{51} & \ding{51} & \ding{55} \\
\cite{26}         & Inverse ML model-based error correction     & RA gain optimization                                    & Not Required & \ding{51} & \ding{51} & \ding{51} & --        & \ding{51} & --        \\
\cite{30}         & Semi-analytical disaggregated GGN model     & QoT estimation                                          & Required     & \ding{55} & \ding{51} & \ding{51} & \ding{51} & --        & \ding{51} \\
\cite{33}         & Enhanced FWM model                           & QoT estimation                                          & Required     & \ding{51} & \ding{51} & \ding{51} & \ding{51} & --        & \ding{51} \\
\cite{10214139}   & DNN-based DT model                          & GSNR estimation \& optimization                         & Not Required & \ding{51} & \ding{51} & \ding{55} & \ding{55} & \ding{51} & \ding{51} \\
\cite{35}         & Insertion loss evaluation-driven GN model   & QoT estimation improvement                               & Required     & \ding{55} & \ding{51} & \ding{51} & \ding{51} & --        & \ding{51} \\
\cite{10526518}   & Closed-form coherent GN model               & QoT estimation                                          & Required     & \ding{51} & \ding{51} & \ding{51} & \ding{51} & --        & \ding{51} \\
\cite{34}         & GN-based model with two-stage power profile & SNR estimation                                          & Required     & \ding{55} & \ding{51} & \ding{51} & \ding{51} & \ding{51} & \ding{51} \\
This work         & Link-adaptive digital twin (LA-DT)          & NLI, ASE and signal power prediction; GSNR estimation   & Not Required & \ding{51} & \ding{51} & \ding{51} & \ding{51} & \ding{51} & \ding{51} \\
\bottomrule
\end{tabularx}
\end{table*}

Table~\ref{tab:relatedworks} summarizes the representative modeling methods in ultra-wideband optical networks, highlighting their modeling objectives, the requirement for accurate link parameter measurement, the online inference capability, and the generalization to diverse link parameters. 

\rrev{In the category of power evolution prediction, Song et al. proposed a physics-informed neural network (PINN) to predict multi-channel power evolution and simultaneously identify fiber attenuation and Raman gain spectrum in C+L-band transmission systems~\cite{20}. Ghodsifar et al. trained gradient-boosting and neural network (NN) regressors on datasets generated by a semi-closed-form GN/EGN model to estimate span-end power and GSNR~\cite{10647400}. Rosa Brusin et al. proposed ANN-based models to accurately predict spectral power profiles under ISRS in ultra-wideband optical networks, supporting real-time QoT estimation with low error across diverse fiber types and channel powers~\cite{21}.}

For RA modeling and optimization in ultra-wideband optical networks, several ML-driven methods have been developed. \rrev{Rosa Brusin et al. presented load-aware neural networks to improve RA gain prediction accuracy under dynamic traffic conditions~\cite{23}. De Moura et al. introduced a NN-based inverse design approach for ultra-wideband RAs, enabling ultra-fast pump configuration to realize arbitrary gain profiles across C+L and S+C+L bands with low errors~\cite{9239870}.  They further proposed a fiber-agnostic NN model for ultra-wideband RAs, combining synthetic datasets and transfer learning to achieve accurate gain prediction across unseen fiber length and types with only a few experimental samples \cite{24}. Liu et al. proposed SMOF, a simultaneous modeling and optimization framework that jointly trains a NN-based digital twin and updates pump configurations online, enabling efficient C+L-band Raman gain optimization with ~90\% less data compared to offline-trained models~\cite{27}. Zhang et al. leveraged parameter-refinement-assisted digital twins to improve prediction robustness under uncertain inputs~\cite{25}.  Gao et al. proposed a fast online optimization method for multi-pump RAs, enabling efficient field RA deployment in ultra-wideband optical netwokrs~\cite{26}.}

\rrev{In the field of QoT estimation, D’Amico et al. proposed a generalized Raman scattering model that accounts for arbitrary loss and Raman scattering, enabling rapid SNR estimation in ultra-wideband optical networks~\cite{30}. Souza et al. proposed the enhanced FWM (eFWM) model for QoT estimation in ultra-wideband optical networks, improving NLI accuracy over the conventional FWM model~\cite{33}. Zhang et al. proposed a digital-twin-assisted GSNR optimization scheme with a symmetric NN-based inverse–forward structure, enabling flexible optical power control in C- and C+L-band systems under dynamic load conditions~\cite {10214139}. Zhou et al. proposed an insertion loss distribution evaluation method based on OCM measurements, improving GN-based QoT estimation accuracy in C+L-band networks under strong SRS~\cite{35}. Zhang et al. proposed a closed-form coherent GN model applicable to arbitrary flexible-grid and heterogeneous links, enabling fast and accurate QoT estimation without resorting to numerical integration~\cite{10526518}. Kimura et al. proposed a two-stage power profile calculation method accounting for Raman pump depletion, enabling accurate GN-based SNR estimation in hybrid EDFA/RA C+L-band networks~\cite{34}.}

In summary, existing physical-layer modeling approaches still fail to jointly satisfy the requirements of high accuracy, low computational complexity, and robust generalization across diverse link parameters. In this paper, we propose the LA-DT that simultaneously achieves real-time inference, low computational complexity, and strong generalization in hybrid-amplified ultra-wideband links. The LA-DT can adapt to diverse link conditions, including variations in fiber length, launch power, Raman pump power, and insertion losses.
Moreover, it explicitly accounts for the variation of Raman pump-induced insertion loss, a factor often overlooked but with a non-negligible impact on network performance.

\section{Physical-layer modeling of hybrid-amplified ultra-wideband networks}
\label{Sec3}
The GSNR is widely recognized as a unified and effective metric for assessing the QoT in ultra-wideband optical networks. This work focuses on analyzing the key factors that contribute to GSNR, with the goal of enabling accurate physical-layer modeling and robust GSNR estimation in diverse link conditions. The GSNR for the $i$-th channel is calculated as:
\begin{equation}
\mathrm{GSNR}_i = \frac{P_{\mathrm{Sig},i}^{\mathrm{Rx}}}{P_{\mathrm{ASE},i}^{\mathrm{Rx}} + P_{\mathrm{NLI},i}^{\mathrm{Rx}}},
\end{equation}
where $P_{Sig,i}^{RX}$, $P_{ASE,i}^{RX}$, and $P_{NLI,i}^{RX}$ denote the signal power, the ASE noise power and the NLI noise power on the $i$-th channel after transmission, respectively.

This work considers the hybrid-amplified conditions with counter-propagating distributed RA. Taking into account ISRS effects, the evolution of signal power for the $i$-th channel along the fiber can be described as follows:
\begin{align}
\frac{\partial P_{Sig,i}}{\partial z} =\,
& - \sum_{k=i+1}^{N_{\mathrm{ch}}} \frac{f_k}{f_i} \, g_i(|f_i - f_k|)\, P_{Sig,k} P_{Sig,i} \notag \\
& + \sum_{k=1}^{i-1} g_i(|f_i - f_k|)\, P_{Sig,k} P_{Sig,i}
+ \sum_{p:\, f_i<f_p} g_i(|f_i - f_p|)\, P_p P_{Sig,i}
- \alpha_i P_{Sig,i}
\end{align}
where $N_{ch}$ is the number of signal channels, $P_{Sig,i}$ and $f_i$ are the power and frequency of the $i$-th signal channel, $P_{Sig,k}$ and $f_k$ are those of the $k$-th signal channel, $P_p$ and $f_p$ are the power and frequency of Raman pumps, and $\alpha_i$ is the fiber attenuation coefficient. $g_i(\cdot)$ denotes the Raman gain coefficient which depends on the frequency separation between the $i$-th signal and the interacting wave.

Typically, the initial value of the signal power at $z=0$ corresponds to the launch power. When considering fiber insertion loss, the signal power should be adjusted by subtracting the input insertion loss. Similarly, at the fiber output, the signal power obtained from the propagation equation should be corrected to account for the output insertion loss. In this work, pump insertion loss is explicitly considered for the first time, which results in the initial pump power being the attenuated value rather than the nominal setting. The received signal power is given by $P_{Sig,i}^{RX} = G_iP_{Sig,i}$, where $G_i$ denotes the gain of the EDFA applied to the $i$-th signal channel at the fiber output.

In hybrid-amplified links, the ASE noise introduced by both RA and EDFA should be taken into account. The total ASE noise power for the $i$-th channel is given by
\begin{equation}
P_{\mathrm{ASE},i}^{\mathrm{Rx}} = P_{\mathrm{ASE},i}^{\mathrm{EDFA}} + G_i P_{\mathrm{ASE},i}^{\mathrm{RA}},
\end{equation}
$P_{ASE,i}^{EDFA}$ represents the ASE noise contribution from the EDFA, calculated as
\begin{equation}
P_{\mathrm{ASE},i}^{\mathrm{EDFA}} = 2 n_{\mathrm{sp},i} (G_i - 1)\, h f_i B_i,
\end{equation}
where $n_{sp,i}$ is the spontaneous emission factor, which is calculated from the amplifier gain $G_i$ and noise figure $NF_i$, and is given by $n_{sp,i} \approx \frac{G_iNF_i-1}{2(G_i-1)}$. $h$ is the Planck constant, $f_i$ and $B_i$ are the frequency and bandwidth of the $i$-th channel.
$P_{ASE,i}^{RA}$ represents the ASE noise power of the $i$-th channel introduced by the distributed RA, which accumulates along the fiber span, and is obtained by solving the following set of coupled differential equations:
\begin{align}
\frac{\partial P_{\mathrm{ASE},i}^{RA}}{\partial z} =\,
& - \sum_{k=i+1}^{N_{\mathrm{ch}}} \frac{f_k}{f_i} g_i(|f_i - f_k|)(P_{Sig,k} + P_{\mathrm{ASE},k}^{RA})
  + \sum_{k=1}^{i-1} g_i(|f_i - f_k|)(P_{Sig,k} + P_{\mathrm{ASE},k}^{RA}) \notag \\
& \quad + \sum_{p:\, f_i < f_p} g_i(|f_i - f_p|)(P_p + P_{\mathrm{ASE},p}^{RA})
  \times \left(P_{\mathrm{ASE},i}^{RA} + 2 h \kappa B_i f_i \right) 
  - \alpha_i P_{\mathrm{ASE},i}^{RA},
\end{align}
with $\kappa = 1+\eta = 1/(1-exp(-h\Delta f/k_B/T))$, where $\eta$ is the phonon occupancy factor, $\Delta f$ denotes the frequency separation between the interacting waves, $K_B$ is the Boltzmann constant and $T$ is the temperature.

The received NLI power for the $i$-th channel can be given by $P_{NLI,i}^{RX} = G_iP_{NLI,i}$, where $P_{NLI,i}$ denotes the accumulated NLI power along the fiber span, prior to the EDFA amplification, and is calculated as:
\begin{equation}
P_{\mathrm{NLI},i} = \eta_n(f_i)\, \left( P_{\mathrm{Sig},i} \right)^3,
\end{equation}
where $\eta_n(f_i)$ is the nonlinear coefficient of the $i$-th channel, \rev{and its closed-form expression has been derived in~\cite{36} based on the GN model extended to incorporate distributed RA and ISRS.} The coefficient is composed of both self-phase modulation (SPM) and cross-phase modulation (XPM) contributions, and can be approximated as:
\begin{equation}
\eta_n(f_i) \approx \sum_{j=1}^{n} \left( \frac{P_{i,j}}{P_i} \right)^2 
\left[ \eta_{\mathrm{SPM},j}(f_i) \cdot n^\epsilon + \eta_{\mathrm{XPM},j}(f_i) \right],
\end{equation}
where $\eta_{\mathrm{SPM},j}(f_i)$ represents the SPM contribution and $\eta_{\mathrm{XPM},j}(f_i)$ represents the total XPM contribution to the NLI within the $j$-th span. $P_{i,j}$ is the power of the $i$-th channel launched into corresponding span, and $\epsilon$ is the coherence factor that characterizes the degree of coherent accumulation of the SPM component across spans.

The GSNR formulation reveals that the modeling of hybrid-amplified ultra-wideband links is highly complex. The signal power evolves under the influence of ISRS and must be calculated by solving a set of coupled ordinary differential equations (ODEs). The ASE noise power, introduced by distributed RA, also requires solving ODEs along the fiber span. Meanwhile,  \rev{the NLI power must be computed through per-channel numerical integration,} which remains computationally intensive even when closed-form expressions are available. Moreover, these closed-form models rely on precise measurements of link parameters, which may not always be accessible in real-world deployments. While machine learning-based methods can alleviate the computational complexity of GSNR modeling, the derived expressions reveal that the GSNR is affected by various link parameters, including EDFA configuration, launch power, fiber length, Raman pump power, and insertion loss. However, conventional machine learning models often suffer from limited generalization capability, making it difficult to maintain high prediction accuracy across different link settings with varying parameter values.

To address the above challenges, we propose the link-adaptive digital twin (LA-DT) to achieve robust physical-layer
modeling in hybrid-amplified ultra-wideband links. Rather than directly modeling the GSNR, we adopt a decomposed modeling strategy that focuses on predicting three key powers. It helps isolate the impact of EDFA configuration variations and facilitates accurate observation and interpretation of individual power parameters under diverse link conditions.

Specifically, we construct three separate DeepModNet-based DT models to predict the NLI noise power, ASE noise power from distributed RA, and signal power before entering the EDFA, and the predicted powers for the $i$-th channel are denoted as $P_{Sig,i}^{Pre}, P_{ASE,i}^{RA,Pre}$, and $P_{NLI, i}^{Pre}$, respectively. The proposed LA-DT and architecture of DeepModNet-based DT models are detailed in Section~\ref{Sec4}. Based on these predicted values, the estimated GSNR for the $i$-th channel is calculated as:
\begin{equation}
\label{eq7}
\mathrm{GSNR}_i^{\mathrm{Est}} = \frac{G_i P_{\mathrm{Sig},i}^{\mathrm{Pre}}}
{\left[ G_i P_{\mathrm{ASE},i}^{\mathrm{RA,Pre}} + 2 n_{\mathrm{sp},i} (G_i - 1)h f_i B_i \right] + G_i P_{\mathrm{NLI},i}^{\mathrm{Pre}}}.
\end{equation}

\begin{figure*}[!t]
\centering
\includegraphics[width=\textwidth]{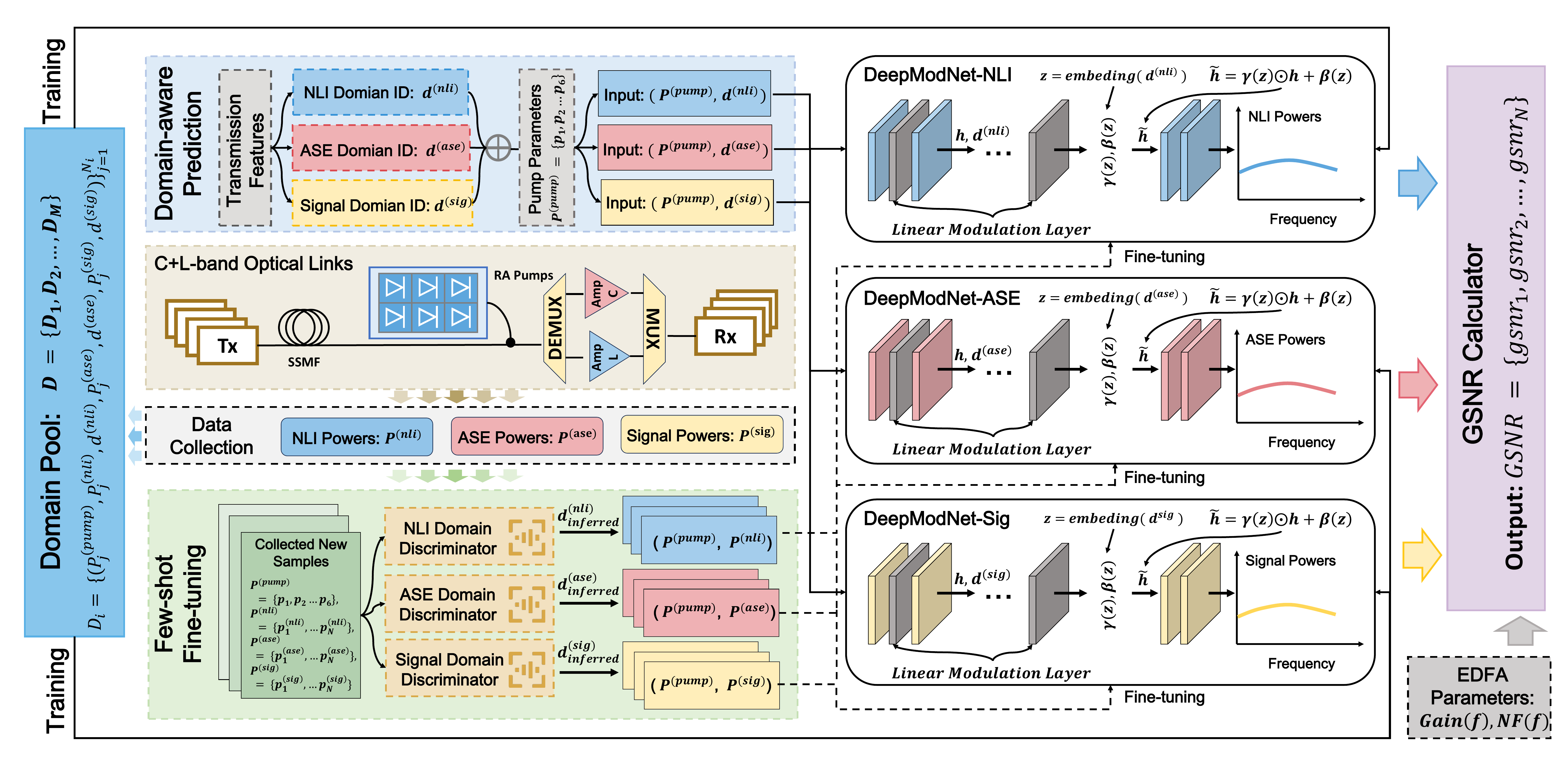}
\caption{The link-adaptive digital twin of hybrid-amplified C+L-band optical links.}
\label{fig1}
\end{figure*}

\section{Link-adaptive digital twin (LA-DT)}
\label{Sec4}
The overall LA-DT is illustrated in Fig.~\ref{fig1}. It comprises three DeepModNet-based DT models that are individually used to predict the NLI, ASE, and signal powers. \rev{Based on these models, LA-DT supports two core operating modes: "domain-aware prediction" and "few-shot fine-tuning". The former is designed to enhance cross-link generalization across known scenarios, while the latter focuses on enabling rapid adaptation to new link conditions. Three DeepModNet-based DT models are pre-trained using a pre-constructed domain pool that covers diverse link conditions. The domain-aware prediction mode directly applies these models by providing the corresponding domain label, enabling accurate power prediction in known scenarios. The few-shot fine-tuning mode is designed around adaptation. When the framework encounters previously unseen scenarios in which link parameters are either unmeasurable or outside the training domain, three domain discriminators are employed to infer the domain label, and model adaptation is carried out by fine-tuning the LMLs using a small number of labeled samples. This section introduces the architecture of DeepModNet, the detailed parameters of three DeepModNet-based DT models, and the generalization and adaptation mechanisms of the two operating modes, explaining how these designs are intended to support fast power prediction and accurate GSNR estimation across diverse transmission scenarios.}

\subsection{DeepModNet: a neural network architecture enhanced by linear modulation layers}
Despite significant differences in link parameter configurations across various optical links, the underlying physical processes, such as signal power evolution, ASE noise accumulation, and nonlinear interference, follow consistent physical laws. These scenarios tend to exhibit similar feature distributions and mapping patterns in modeling tasks. As a result, the overall mapping patterns are transferable, while the specific functional relationships may shift due to variations in link conditions. To exploit this property, we propose DeepModNet, which integrates LMLs into the neural network for physical-layer modeling, inspired by the concept of feature-wise conditioning~\cite{37}. This mechanism enables dynamic conditional modeling across domains by modulating internal features based on domain labels. It allows the model to flexibly adapt to domain-specific characteristics across different link conditions, while maintaining effective parameter sharing within the backbone network across domains.

\begin{table}[h]
\scriptsize
\centering
\caption{Link configuration parameters of 35 scenarios in domain pool}
\label{tab1}
\begin{tabularx}{\linewidth}{>{\centering\arraybackslash}X
                        >{\centering\arraybackslash}X
                        >{\centering\arraybackslash}X
                        >{\centering\arraybackslash}X
                        >{\centering\arraybackslash}X}
\toprule
\textbf{Domain Index} & \textbf{Fiber Length (km)} & \textbf{Launch Power (dBm)} & \textbf{Pump Insertion Loss (dB)} & \textbf{Fiber Insertion Loss (dB)} \\
\midrule
1  & 70 &  0  & 0    & 0.5 \\
2  & 40 &  0  & 0    & 0.5 \\
3  & 50 &  0  & 0    & 0.5 \\
4  & 60 &  0  & 0    & 0.5 \\
5  & 80 &  0  & 0    & 0.5 \\
6  & 70 & -3  & 0    & 0.5 \\
7  & 70 & -2  & 0    & 0.5 \\
8  & 70 & -1  & 0    & 0.5 \\
9  & 70 &  1  & 0    & 0.5 \\
10 & 70 &  2  & 0    & 0.5 \\
11 & 50 & -3  & 0    & 0.5 \\
12 & 50 & -2  & 0    & 0.5 \\
13 & 50 & -1  & 0    & 0.5 \\
14 & 50 &  1  & 0    & 0.5 \\
15 & 50 &  2  & 0    & 0.5 \\
16 & 40 &  0  & 0.5  & 0.5 \\
17 & 50 &  0  & 0.5  & 0.5 \\
18 & 60 &  0  & 0.5  & 0.5 \\
19 & 70 &  0  & 0.5  & 0.5 \\
20 & 80 &  0  & 0.5  & 0.5 \\
21 & 40 &  0  & 0    & 0.75 \\
22 & 50 &  0  & 0    & 0.75 \\
23 & 60 &  0  & 0    & 0.75 \\
24 & 70 &  0  & 0    & 0.75 \\
25 & 80 &  0  & 0    & 0.75 \\
26 & 40 &  0  & 0    & 0.25 \\
27 & 50 &  0  & 0    & 0.25 \\
28 & 60 &  0  & 0    & 0.25 \\
29 & 70 &  0  & 0    & 0.25 \\
30 & 80 &  0  & 0    & 0.25 \\
31 & 40 &  0  & 0.25 & 0.5 \\
32 & 50 &  0  & 0.25 & 0.5 \\
33 & 60 &  0  & 0.25 & 0.5 \\
34 & 70 &  0  & 0.25 & 0.5 \\
35 & 80 &  0  & 0.25 & 0.5 \\
\bottomrule
\end{tabularx}
\end{table}

\begin{table*}[!h]
\centering
\caption{Training settings and architecture of DeepModNet-based and BaseCondNet-based DT models}
\label{tab2}
\resizebox{\textwidth}{!}{%
\begin{tabular}{lccccccccc}
\toprule
\textbf{Model} & \textbf{Input Dim} & \textbf{Output Dim} & \textbf{Hidden Layers} & \textbf{Hidden Neurons} & \textbf{LMLs} & \textbf{Domain Embedding} & \textbf{Activation} & \textbf{Learning Rate} & \textbf{Optimizer} \\
\midrule
DeepModNet-NLI  & 6  & 120 & 3 & 336  & 1 & 32 & Tanh & 0.001 & Adam \\
DeepModNet-ASE  & 6  & 120 & 4 & 512  & 4 & 32 & Tanh & 0.001 & Adam \\
DeepModNet-Sig  & 6  & 120 & 4 & 2528 & 4 & 64 & ReLU & 0.001 & Adam \\
BaseCondNet-NLI & 10 & 120 & 3 & 432  & -- & -- & Tanh & 0.001 & Adam \\
BaseCondNet-ASE & 10 & 120 & 4 & 768  & -- & -- & Tanh & 0.001 & Adam \\
BaseCondNet-Sig & 10 & 120 & 4 & 1024 & -- & -- & ReLU & 0.001 & Adam \\
\bottomrule
\end{tabular}%
}
\end{table*}

Specifically, each LML performs dynamic conditioning by learning domain-dependent scaling and shifting functions. Given an intermediate feature vector $h$ and an embedded domain label $z$, the LML applies the following transformation to obtain the modulated feature $\tilde{h}$:
\begin{equation}
\tilde{h} = \gamma(z) \odot h + \beta(z),
\end{equation}
where $\gamma(z)$ and $\beta(z)$ are learnable functions that generate modulation parameters conditioned on the domain $z$, and $\odot$ denotes element-wise multiplication. This design allows the model to adaptively modulate internal representations in a domain-aware manner without altering the shared architecture, achieving a practical trade-off between specificity and parameter efficiency. As a result, DeepModNet achieves robust generalization in diverse scenarios and reduces the dependence on extensive training datasets.

\subsection{Domain-aware power prediction via DeepModNet-based DT model}
We construct a comprehensive domain pool comprising 35 transmission scenarios with diverse ultra-wideband link conditions, including variations in fiber length, per-channel launch power, fiber insertion loss, and Raman pump insertion loss, as summarized in Table ~\ref{tab1}. \rev{We explicitly consider insertion loss variations at both ends of the fiber. The transmitter-side insertion loss affects the launched signal powers, thereby altering the nonlinear power evolution along the span, while the receiver-side insertion loss introduces an additional linear attenuation at the end of the link. The superposition of these two effects results in a combined impact that differs from a simple launch power variation and cannot be represented by a single scaling factor.} The considered condition is C+L band system with a total bandwidth of 12 THz, divided into 120 WDM channels at 100 GHz spacing. Standard single-mode fiber (SSMF) is employed with an attenuation of 0.2 dB/km, dispersion of 16.7 ps/nm/km at 1550 nm, and dispersion slope of 0.08 ps/nm²/km. In the modeling setup, the six-dimensional Raman pump powers are treated as input features to the model, with six pumps employed at wavelengths of 1425 nm, 1435 nm, 1445 nm, 1465 nm, 1480 nm, and 1500 nm, each randomly sampled from 0 to 200 mW with 1 mW steps. The domain-specific parameters described in Table ~\ref{tab1} are embedded into the model and used to guide the feature modulation process. For each data sample, the output includes the NLI power, ASE power, signal power and the GSNR, all measured across 120 channels. \rev{The training data are generated using GNPy~\cite{38}, which adopts a semi-analytical GN/EGN framework for physical layer modeling. In this framework, Raman amplification is modeled through channel-dependent Raman gain coefficients and numerical solutions of the coupled power evolution equations. This approach captures distributed Raman gain and ISRS-induced power transfer along the fiber. Only the gain-related component of the Raman response, corresponding to the imaginary part of the Raman susceptibility, is considered. The triangle approximation of the Raman gain spectrum is not assumed in this work.} We collect 500 samples per scenario, with 300 used for training and 200 for testing, forming a multi-domain dataset comprising 10,500 training samples and 7,000 test samples.

Three DT models targeting different types of power prediction are constructed, namely DeepModNet-NLI, DeepModNet-ASE, and DeepModNet-Sig, which are used to predict NLI power, ASE noise power introduced by distributed RA, and signal power, respectively. These models are trained on the constructed multi-domain training dataset. During training, each model adopts the mean squared error (MSE) as the loss function for the regression task, which is defined as follows:
\begin{equation}
\mathcal{L}_{\text{MSE}} = \frac{1}{N} \sum_{i=1}^{N} \left\| \hat{\mathbf{y}}_i - \mathbf{y}_i \right\|_2^2,
\end{equation}
where $\hat{\mathbf{y}}_i$ denotes the predicted output vector of the $i$-th sample, $\mathbf{y}_i$ is the corresponding ground-truth label, and $N$ is the total number of training samples. The training settings and finalized model architectures are summarized in Table~\ref{tab2}. Specifically, all three models take as input a 6-dimensional vector representing the Raman pump powers, and produce a 120-dimensional output corresponding to the 120 WDM channels in the C+L band scenario. DeepModNet-NLI consists of 3 hidden layers with a total of 336 neurons, employs the Tanh activation function, and incorporates 1 LML along with a 32-dimensional domain embedding. Both DeepModNet-ASE and DeepModNet-Sig employ 4 hidden layers. The former uses a total of 512 neurons with Tanh activation, while the latter uses  a total of 2528 neurons with ReLU activation. Each of them contains 4 LMLs, with domain embeddings of 32 and 64 dimensions, respectively. All models are trained using the Adam optimizer with a learning rate of 0.001.

We adopt a static conditional modeling approach as the baseline to construct comparison DT models, referred to as BaseCondNet, which is widely used in generalized modeling tasks. Instead of applying dynamic feature modulation, the baseline method directly incorporates scenario-specific parameters into the input features. In the setting, the model input consists of 10 features, including a six-dimensional vector representing the Raman pump powers and four link parameters: fiber length, per-channel launch power, fiber insertion loss, and Raman pump insertion loss. BaseCondNet is trained using the same dataset and loss function as DeepModNet, and three models, BaseCondNet-NLI, BaseCondNet-ASE, and BaseCondNet-Sig, are constructed accordingly. Unlike DeepModNet, BaseCondNet does not incorporate domain embedding or LMLs. The detailed training settings and model architectures are summarized in Table~\ref{tab2}.

To evaluate the modeling performance of the DT models, two metrics are adopted in this paper: 1) power prediction accuracy, measured by the error between the predicted and actual power values across all 120 WDM channels; 2) GSNR estimation accuracy, where the estimated GSNR is calculated for each channel based on the predicted NLI power, ASE power and signal power, as defined in Eq.~(\ref{eq7}). The RMSE between the estimated and actual GSNR values serves as the performance indicator for the DT models.

\subsection{Few-shot fine-tuning via domain discriminator}
\rev{In practical applications, the constructed DeepModNet-based DT models may encounter two types of challenging scenarios: one involves previously unseen configurations that are not covered in the training dataset, and the other involves scenarios where link parameters have changed but the updated parameter values are unknown or unmeasurable. In such cases, the models are required to rapidly adapt to these new link conditions in order to maintain accurate power prediction and reliable GSNR estimation. However, explicit domain labels are difficult to obtain in both situations, which makes the fine-tuning process of the three DT models infeasible, as it relies on such information.}

To address this issue, we design three domain discriminators, each corresponding to one of the DT models (DeepModNet-NLI, DeepModNet-ASE, and DeepModNet-Sig). Unlike conventional domain identification methods based on input feature similarity or distributional measures such as maximum mean discrepancy (MMD),  the proposed domain discriminators perform domain identification by evaluating the consistency of functional mappings. The underlying principle is to select the training domain whose functional mapping exhibits the highest consistency with that of the target samples under the current model architecture. 

Specifically, given the input set of $N$ target samples $\mathcal{X}_{\text{tgt}} = \Big\{ \mathbf{x}^{(i)}_{\text{tgt}} \Big\}_{i=1}^N$, and the corresponding output set $\mathcal{Y}_{\text{tgt}} = \Big\{ \mathbf{y}^{(i)}_{\text{tgt}} \Big\}_{i=1}^N$. The domain discriminator selects the optimal domain label $d^*$ from the domain pool $\mathcal{D}_{\text{pool}}$ for the target samples. This selection is achieved by minimizing the following objective function:
\begin{equation}
d^* = \arg\min_{d \in \mathcal{D}_{\text{pool}}} \frac{1}{N} \sum_{i=1}^{N} \left\| f_d\left( \mathbf{x}^{(i)}_{\text{tgt}} \right) - \mathbf{y}^{(i)}_{\text{tgt}} \right\|_2^2,
\end{equation}
where $f_d(\cdot)$ denotes the mapping function of the pre-trained model under domain $d$, and $\mathbf{x}^{(i)}_{\text{tgt}}$ and $\mathbf{y}^{(i)}_{\text{tgt}}$ represent the input and output of the $i$-th target sample, respectively. It is applicable to three DeepModNet-based DT models and enables few-shot fine-tuning in scenarios where domain labels are unavailable.

After identifying the optimal training domain, the model assigns the corresponding domain labels to the target samples to guide the modulation of LMLs. Few-shot fine-tuning is then performed, during which the backbone network remains frozen and only the parameters of LMLs are updated using a small number of labeled target samples. Optimization is conducted with the previously defined MSE loss. Rapid adaptation to the target scenario is achieved through this fine-tuning process.

To evaluate the effectiveness of the few-shot fine-tuning mechanism guided by the domain discriminator, we follow the evaluation method described previously and adopt two performance metrics: 1) power prediction accuracy, measured by the error between predicted and actual power values across 120 channels; and 2) GSNR estimation accuracy, obtained by  computing the RMSE between the actual GSNR and the values derived by substituting the fine-tuned predictions of the three powers into Eq.~(\ref{eq7}). These metrics reflect the performance of DeepModNet-based DT models in terms of their rapid adaptability and modeling accuracy under scenarios with previously unseen configurations or missing link parameters.

\section{Results and Discussions}
\label{Sec5}
\subsection{Domain-aware prediction: DeepModNet $vs$ BaseCondNet}

The performance of the developed DeepModNet-based DT models is compared with that of the baseline BaseCondNet-based models across the 35 ultra-wideband scenarios listed in Table~\ref{tab1}. \rev{It should be noted that the scenarios considered in this work are limited to equivalent-span configurations, and the current model does not explicitly support heterogeneous multi-span links. Within this scope, the proposed LA-DT framework captures link-level physical parameter variations, including variations in EDFA gain and noise figure, Raman pump powers, fiber length, launch signal power, fiber insertion loss, and pump insertion loss.} A total of six models, three based on DeepModNet and three based on the baseline BaseCondNet, are evaluated under each scenario, respectively targeting the prediction of NLI power, ASE power, and signal power. Each scenario includes 200 independent test samples, resulting in a total of 7,000 test cases,  all of which are strictly separated from the training data.

\begin{figure*}[!t]
\centering
\includegraphics[width=\textwidth]{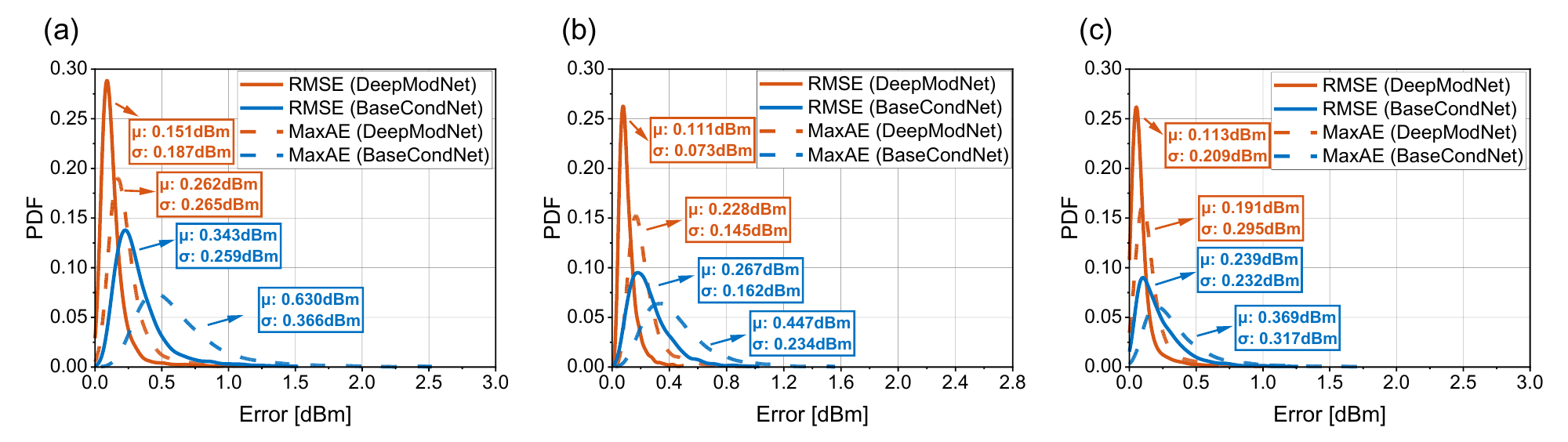}
\caption{Probability density function (PDF) distributions of RMSE and MaxAE for DeepModNet-based and BaseCondNet-based DT models in the prediction of: (a) NLI power, (b) ASE power, and (c) signal power.}
\label{fig2}
\end{figure*}

Model performance is evaluated based on prediction errors across 120 WDM channels, using two metrics: root mean squared error (RMSE) and maximum absolute error (MaxAE). RMSE reflects the overall prediction accuracy, while MaxAE captures the largest per-channel deviation within each sample. The corresponding results are illustrated in Fig.~\ref{fig2}, where Fig.~\ref{fig2}(a) presents the error distribution for the NLI power prediction task, and Fig.~\ref{fig2}(b) and Fig.~\ref{fig2}(c) show results for the ASE power and signal power modeling tasks, respectively. All distributions are visualized in the form of probability density functions (PDFs), allowing a clear comparison of modeling accuracy between two approaches across all prediction tasks.

As shown in Fig.~\ref{fig2}, DeepModNet-based DT models consistently outperform the baseline BaseCondNet-based models across all three power prediction tasks, despite being trained with the same amount of data. For the NLI power prediction task, the mean and standard deviation ($\mu \pm \sigma$) of the RMSE are $0.151 \pm 0.187~\mathrm{dBm}$ for DeepModNet-NLI and $0.343 \pm 0.259~\mathrm{dBm}$ for BaseCondNet-NLI. The corresponding MaxAE values are $0.262 \pm 0.265~\mathrm{dBm}$ and $0.630 \pm 0.366~\mathrm{dBm}$, respectively. For the ASE power prediction task, the RMSE values are $0.111 \pm 0.073~\mathrm{dBm}$ for DeepModNet-ASE and $0.267 \pm 0.162~\mathrm{dBm}$ for BaseCondNet-ASE. The corresponding MaxAE values are $0.228 \pm 0.145~\mathrm{dBm}$ and $0.447 \pm 0.234~\mathrm{dBm}$, respectively. For the signal power prediction task, the RMSE values are $0.113 \pm 0.209~\mathrm{dBm}$ for DeepModNet-Sig and $0.239 \pm 0.232~\mathrm{dBm}$ for BaseCondNet-Sig. The corresponding MaxAE values are $0.191 \pm 0.295~\mathrm{dBm}$ and $0.369 \pm 0.317~\mathrm{dBm}$, respectively.

By taking the average RMSE over all 7,000 test samples as the primary indicator of prediction accuracy, significant improvements are achieved by DeepModNet over BaseCondNet under identical training data conditions. Specifically, DeepModNet-NLI improves the prediction accuracy by 56.0\% compared to BaseCondNet-NLI in the NLI power modeling task. Similarly, DeepModNet-ASE and DeepModNet-Sig achieve accuracy improvements of 58.4\% and 52.7\%, respectively, in the ASE and signal power prediction tasks.

\begin{figure*}[!t]
\centering
\includegraphics[width=\textwidth]{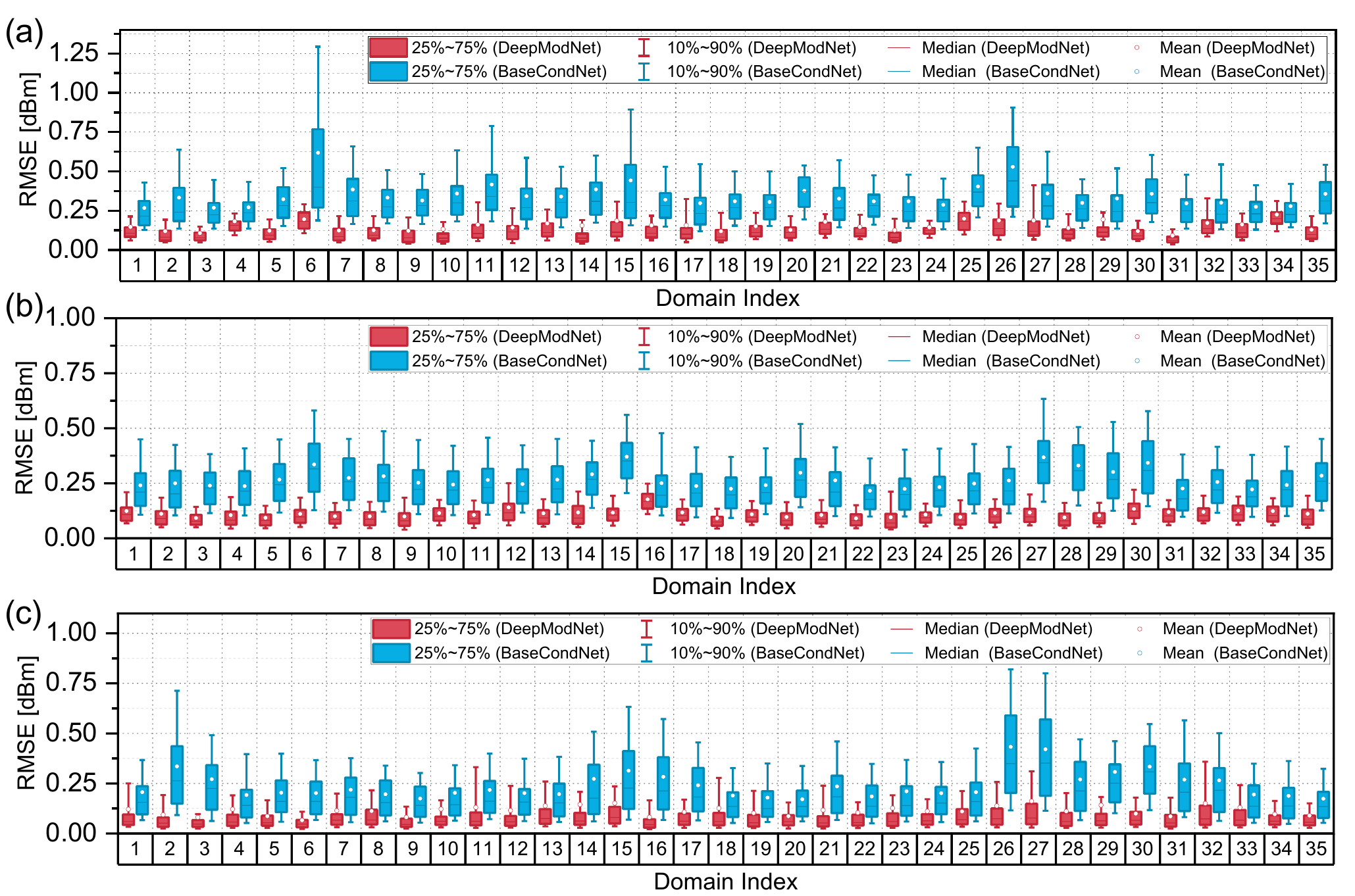}
\caption{Domain-wise box plot distributions of RMSE for DeepModNet-based and BaseCondNet-based DT models in the prediction of: (a) NLI power, (b) ASE power, and (c) signal power.}
\label{fig3}
\end{figure*}

To further analyze the domain-wise modeling performance, we present in Fig.~\ref{fig3} the RMSE distributions of DeepModNet-based and BaseCondNet-based DT models across all 35 test scenarios, with each domain containing 200 test samples. Fig.~\ref{fig3} (a), (b), and (c) correspond to the NLI, ASE, and signal power prediction tasks, respectively. Each box plot summarizes the distribution of RMSE values within a domain: the box spans the interquartile range (25\%$\sim$75\%), the whiskers extend to the 10\% and 90\% percentiles, the horizontal line inside the box represents the median, and the dot denotes the mean. Across all three subplots in Fig.~\ref{fig3}, the red boxes representing DeepModNet-based models are consistently positioned below the blue boxes of BaseCondNet-based models. This consistent visual pattern, together with the lower median RMSE values, smaller interquartile ranges, shorter whiskers, and lower mean RMSE values observed for DeepModNet-based models, indicates a clear performance advantage across all domains. These combined characteristics demonstrate that DeepModNet-based DT models achieve not only lower prediction errors but also more stable and concentrated error distributions under diverse link conditions.

Overall, compared to the baseline conditional modeling approach, the proposed DeepModNet architecture achieves significantly higher prediction accuracy under the same amount of training data. This performance advantage is consistently observed across all three power modeling tasks, with substantial reductions in both RMSE and MaxAE. Furthermore, domain-wise evaluations across all 35 scenarios confirm that DeepModNet-based DT models consistently outperform the baseline models in every individual link condition, demonstrating both stable and robust prediction performance across diverse domains. The observed accuracy improvement is primarily attributed to the dynamic conditional modeling strategy adopted by DeepModNet architecture. Unlike the static modeling mechanism of BaseCondNet, which concatenates link parameters as fixed inputs and relies heavily on large-scale and well-covered training data, DeepModNet introduces LMLs that apply scaling and shifting operations to intermediate features conditioned on domain labels. This modulation mechanism enables the model to effectively capture variations induced by different link configurations, achieving dynamic adjustment and adaptive modeling across diverse transmission scenarios. While enhancing modeling flexibility, this mechanism also preserves the parameter-sharing structure of the backbone network, allowing effective information sharing across different scenarios. As a result, the model reduces its dependence on large-scale domain-specific data and maintains high prediction accuracy even under limited training conditions.

\begin{table}[!h]
\scriptsize  
\centering
\caption{\rev{Runtime comparison between GNPy propagation simulation and DeepModNet-based model inference}}
\label{tab4}
\begin{tabularx}{\linewidth}{>{\centering\arraybackslash}X
                        >{\centering\arraybackslash}X
                        >{\centering\arraybackslash}X
                        >{\centering\arraybackslash}X
                        >{\centering\arraybackslash}X}
\toprule
\textbf{Domain Index} & \textbf{GNPy / 20 runs (s)} & \textbf{DeepModNet-NLI / 20 runs (s)} & \textbf{DeepModNet-ASE / 20 runs (s)} & \textbf{DeepModNet-Sig / 20 runs (s)} \\
\midrule
1  & 8.1486 & 0.0079 & 0.0829 & 0.0499 \\
2  & 7.9300 & 0.0032 & 0.0732 & 0.0412 \\
3  & 8.1129 & 0.0120 & 0.1010 & 0.0414 \\
4  & 8.6278 & 0.0046 & 0.0809 & 0.0430 \\
5  & 8.2477 & 0.0043 & 0.1207 & 0.0580 \\
\bottomrule
\end{tabularx}
\end{table}

\rev{To evaluate the computational efficiency of the proposed framework, we conduct a wall-clock runtime comparison between GNPy propagation simulations and the trained DeepModNet-based models under identical link configurations. Specifically, we select the first five representative domains from the domain pool, which cover different fiber lengths, and measure the execution time in each scenario. For each domain configuration, 20 propagation simulations are performed using GNPy to obtain the received power distributions. Under the same link conditions, 20 forward inference runs are executed for each of the three trained models. The corresponding runtime results are summarized in Table~\ref{tab4}. As shown in Table~\ref{tab4}, GNPy requires approximately 8 seconds to complete 20 runs for each domain configuration. In contrast, the trained DeepModNet-based models complete 20 inference runs within milliseconds. Across all five domains, the total runtime of the ML-based models is substantially lower than that of GNPy, demonstrating the computational advantage of the proposed approach for repeated evaluations. All time measurements are conducted on a local computing system equipped with a 12th Gen Intel Core i7-1260P CPU running at 2.10 GHz and 32 GB of RAM.}

\begin{table}[!b]
\scriptsize  
\centering
\caption{Link configuration parameters of 12 unseen scenarios}
\label{tab3}
\begin{tabularx}{\linewidth}{>{\centering\arraybackslash}X
                        >{\centering\arraybackslash}X
                        >{\centering\arraybackslash}X
                        >{\centering\arraybackslash}X
                        >{\centering\arraybackslash}X}
\toprule
\textbf{Domain Index} & \textbf{Fiber Length (km)} & \textbf{Launch Power (dBm)} & \textbf{Pump Insertion Loss (dB)} & \textbf{Fiber Insertion Loss (dB)} \\
\midrule
1  & 45 & -2.5 & 0.0 & 0.4 \\
2  & 55 & -2.5 & 0.0 & 0.4 \\
3  & 65 & -2.5 & 0.0 & 0.4 \\
4  & 75 & -2.5 & 0.0 & 0.4 \\
5  & 45 & -1.5 & 0.2 & 0.7 \\
6  & 55 & -1.5 & 0.2 & 0.7 \\
7  & 65 & -1.5 & 0.3 & 0.7 \\
8  & 75 & -1.5 & 0.3 & 0.7 \\
9  & 45 &  1.5 & 0.4 & 0.9 \\
10 & 55 &  1.5 & 0.4 & 0.9 \\
11 & 65 &  1.5 & 0.6 & 0.9 \\
12 & 75 &  1.5 & 0.6 & 0.9 \\
\bottomrule
\end{tabularx}
\end{table}

\begin{figure*}[!t]
\centering
\includegraphics[width= \linewidth]{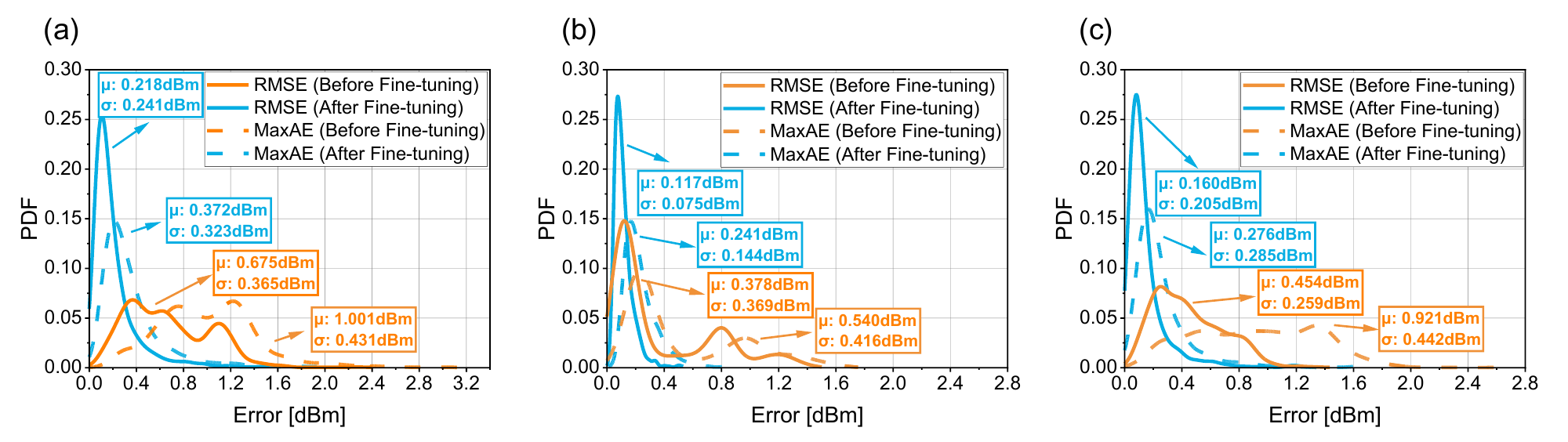}
\caption{Probability density function (PDF) distributions of RMSE and MaxAE for DeepModNet-based DT models before and after fine-tuning in the prediction of: (a) NLI power, (b) ASE power, and (c) signal power.}
\label{fig4}
\end{figure*}

\subsection{Few-shot fine-tuning in unseen scenarios}
To assess the adaptation capability of the domain discriminator–guided DT models to previously unseen scenarios, we construct 12 new ultra-wideband scenarios outside the training domain pool. The link configurations of these scenarios are detailed in Table~\ref{tab3}, covering variations in fiber length, per-channel launch power, and loss characteristics. For each scenario, 20 samples are collected for fine-tuning and 200 samples for testing, resulting in a total of 2,400 independent test samples. 

First, three DeepModNet-based DT models trained during the initial phase without any fine-tuning are directly applied to the 12 new scenarios to establish baseline performance. Then, few-shot fine-tuning is performed using 20 collected samples from each scenario. In this process, designed domain discriminators are used to infer the domain label of new samples to activate corresponding linear modulation parameters. Subsequently, the backbone network is kept frozen, and only the scaling and shifting parameters within the LML corresponding to the inferred domain are updated, enabling rapid adaptation to new scenarios. After fine-tuning, models are re-evaluated on the complete set of 2,400 test samples across all 12 new scenarios.

Fig.~\ref{fig4} presents the PDF distributions of prediction errors before and after few-shot fine-tuning with 20 samples per scenario. Fig.~\ref{fig4} (a), (b), and (c) correspond to the prediction tasks of NLI power, ASE power, and signal power, respectively, evaluated using RMSE and MaxAE metrics. In the NLI power prediction task, the mean and standard deviation ($\mu \pm \sigma$) of RMSE and MaxAE for DeepModNet-NLI before fine-tuning are $0.675 \pm 0.365~\mathrm{dBm}$ and $1.001 \pm 0.431~\mathrm{dBm}$, respectively. After fine-tuning with 20 samples, these values are reduced to $0.218 \pm 0.241~\mathrm{dBm}$ and $0.372 \pm 0.323~\mathrm{dBm}$, respectively. In the ASE power prediction task, the RMSE and MaxAE values for DeepModNet-ASE before fine-tuning are $0.378 \pm 0.266~\mathrm{dBm}$ and $0.540 \pm 0.416~\mathrm{dBm}$, respectively. After fine-tuning, they decrease to $0.117 \pm 0.075~\mathrm{dBm}$ and $0.241 \pm 0.144~\mathrm{dBm}$, respectively. In the signal power prediction task, the RMSE and MaxAE values for DeepModNet-Sig before fine-tuning are $0.454 \pm 0.259~\mathrm{dBm}$ and $0.921 \pm 0.442~\mathrm{dBm}$, respectively. After fine-tuning, they are reduced to $0.160 \pm 0.205~\mathrm{dBm}$ and $0.276 \pm 0.285~\mathrm{dBm}$, respectively.

Results demonstrate that with only 20 samples per scenario for fine-tuning, DeepModNet-based DT models achieve a substantial improvement in modeling accuracy under previously unseen conditions. Specifically, based on the average RMSE across 2,400 test samples, the modeling accuracy is improved by 67.7\%, 69.0\%, and 64.8\% for the NLI, ASE, and signal power prediction tasks, respectively. These results validate the effectiveness of the proposed domain discriminator in identifying new scenarios and confirm that the DeepModNet-based DT models possess strong cross-domain generalization capability, enabling rapid adaptation to unseen scenarios even with extremely limited data.

\begin{figure*}[!h]
\centering
\includegraphics[width= \textwidth]{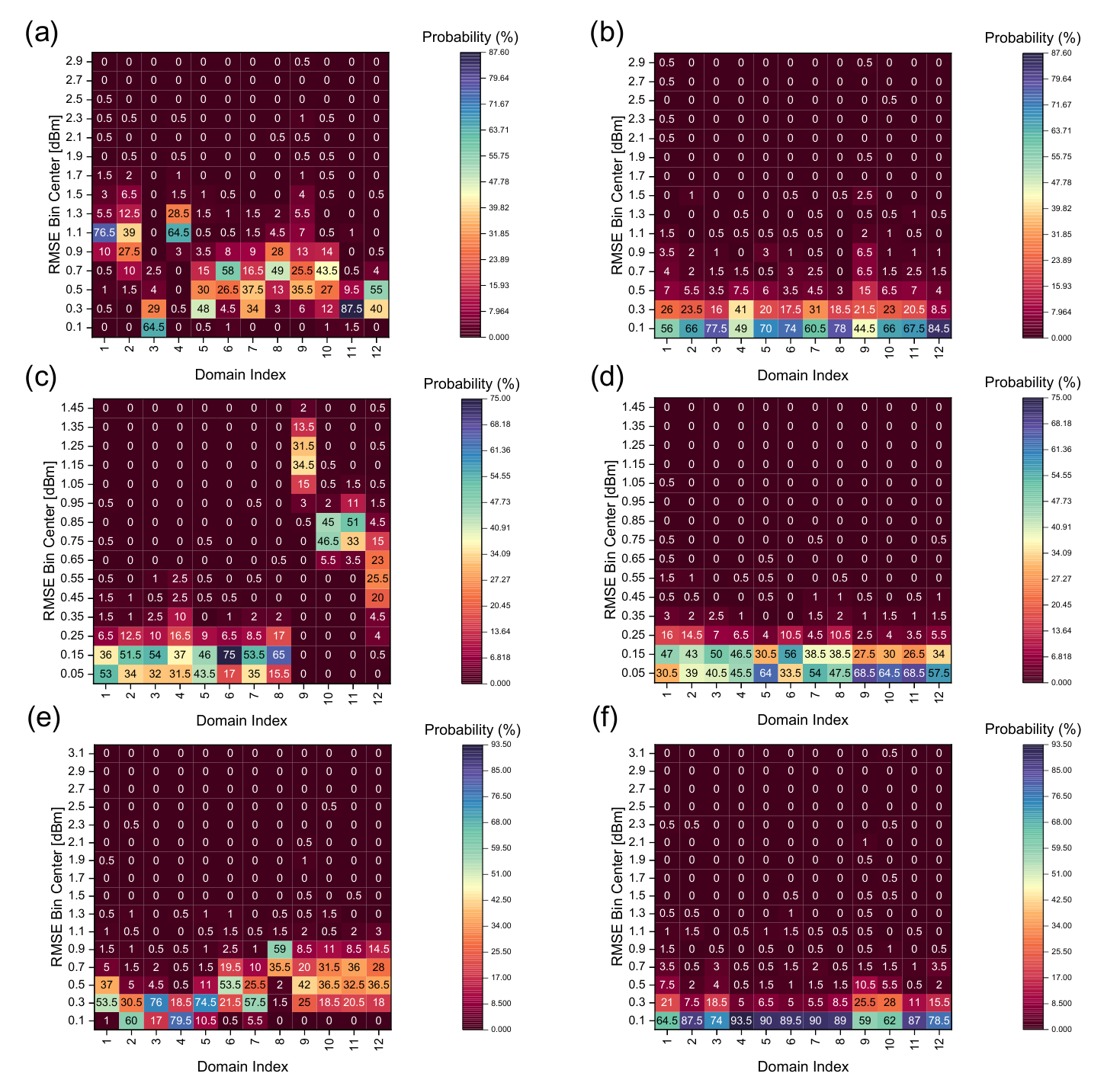}
\caption{Domain-wise heatmap distributions of RMSE for DeepModNet-based DT models before and after fine-tuning in the prediction of: (a) NLI power (before), (b) NLI power (after), (c) ASE power (before), (d) ASE power (after), (e) signal power (before), (f) signal power (after).}
\label{fig5}
\end{figure*}

To further analyze the prediction performance of DeepModNet-based DT models before and after fine-tuning in each new domain, we present the RMSE distribution heatmaps across 12 unseen scenarios in Fig. \ref{fig5}. The vertical axis represents the center values of RMSE intervals, the horizontal axis corresponds to domain index, and the color intensity indicates the probability density (in percentage) associated with each error interval. Fig. \ref{fig5}(a) and Fig. \ref{fig5}(b) show the RMSE distributions of DeepModNet-NLI before and after fine-tuning, respectively; Fig. \ref{fig5}(c) and Fig. \ref{fig5}(d) correspond to DeepModNet-ASE; and Fig. \ref{fig5}(e) and Fig. \ref{fig5}(f) illustrate the results for DeepModNet-Sig.

As shown in Fig.~\ref{fig5}, the error distributions exhibit notable improvements after fine-tuning across all domains. RMSE values tend to concentrate in lower ranges, and the distributions become more compact, with color intensities shifting downward. For DeepModNet-NLI, low-error regions (e.g., $< 0.5~\mathrm{dBm}$) dominate in most domains after fine-tuning, resulting in  a more concentrated heatmap. Similar trends are observed for DeepModNet-ASE and DeepModNet-Sig, where the probabilities of high-error regions significantly decrease after fine-tuning. A closer examination of individual domains reveals that RMSE decreases to varying degrees across all 12 new scenarios, with notably more concentrated and lower error distributions. These results suggest that the model effectively adapts to each unseen domain, even under limited-data conditions. In a few specific domains, such as $Domain 1$ across all three tasks, larger errors are observed after fine-tuning in some cases. In these domains, certain test samples exhibit significant differences from the 20 fine-tuning samples. As the model is updated based only on this limited data, its predictions tend to align with the fine-tuning set, leading to larger prediction errors for those biased test samples. However, results indicate that these exceptions are rare, with all models demonstrating strong performance in most cases and maintaining general accuracy.

Overall, these results further confirm the strong adaptability of DeepModNet-based DT models under limited-data conditions. It consistently achieves high modeling accuracy across diverse unseen scenarios, demonstrating both robustness and strong generalization under data-scarce and unfamiliar conditions.

\subsection{GSNR estimation}
In this section, we evaluate the GSNR estimation performance of the proposed LA-DT under the two operating modes described earlier: one based on domain-aware prediction using fully trained models for known scenarios, and the other based on few-shot fine-tuning for previously unseen domains. Specifically, the predicted values of NLI power, ASE power and signal power are substituted into Eq.~(\ref{eq7}) to compute the estimated GSNR, which is then compared with the ground-truth GSNR. To assess estimation accuracy, we adopt two metrics: root mean squared error (RMSE) and maximum absolute error (MaxAE), calculated across all 120 channels.

As previously described, this work adopts a decomposed modeling approach for GSNR estimation, which enables flexible adaptation to EDFAs with varying configurations. Therefore, in the GSNR estimation process, we consider a variety of EDFA configurations in order to adapt to its heterogeneity in practical optical networks. For each test case, the gain of EDFA is configured based on the fiber attenuation of the corresponding transmission link. All scenarios are based on SSMF and fiber lengths ranging from 40 km to 80 km across different test cases. For the 60 channels in the L-band, the gain of EDFA is set to compensate for the  fiber attenuation, resulting in a gain range of 8 dB to 16 dB. In the C-band, the EDFA gain is increased by an additional 2 dB to compensate for the power transfer caused by ISRS, resulting in a gain range of 10 dB to 18 dB. The noise figure is set to 5 for C-band EDFAs and 6 for L-band EDFAs.

\begin{figure}[t]
\centering
\includegraphics[width= 0.5 \linewidth]{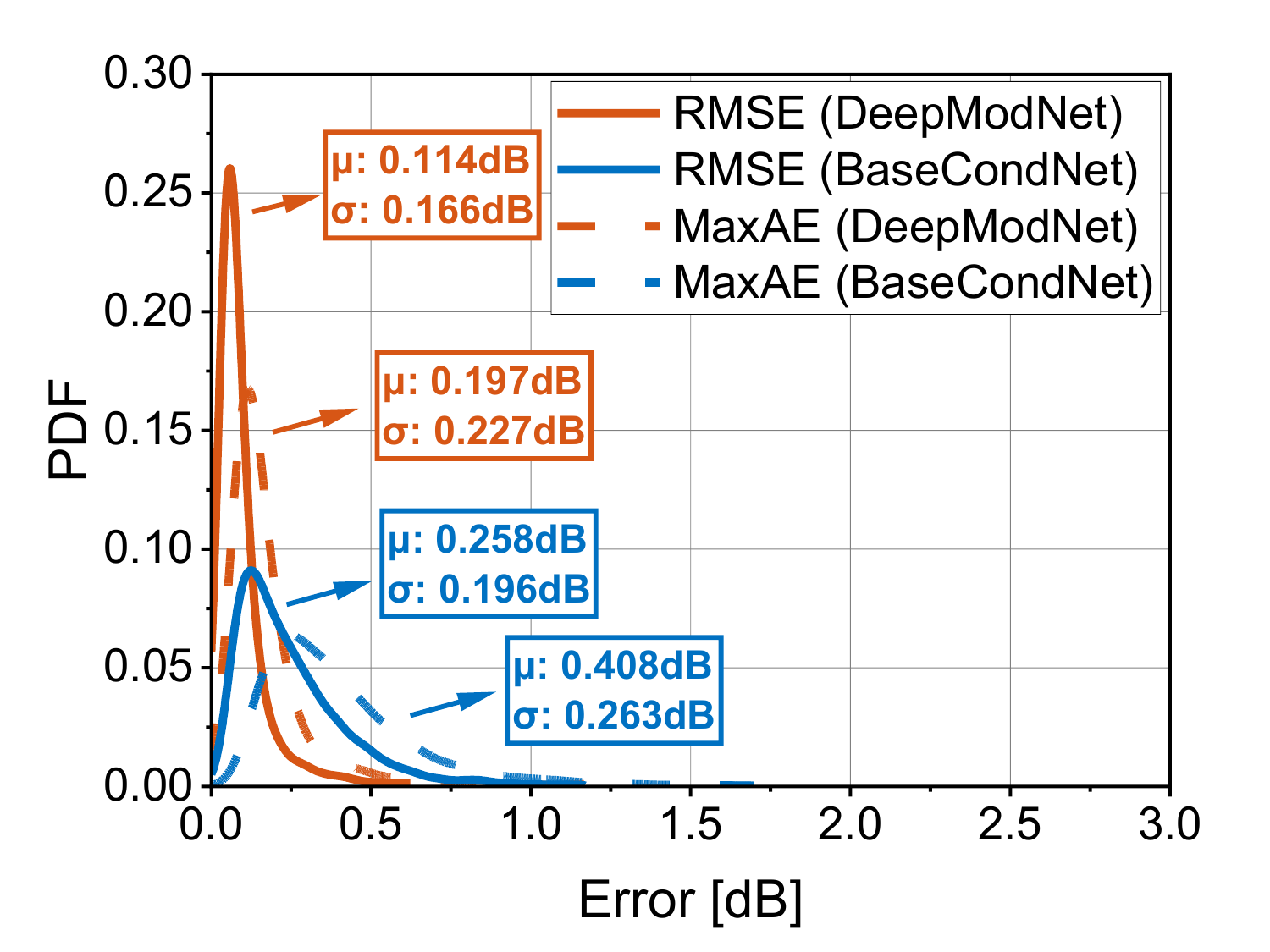}
\caption{Probability density function (PDF) distributions of RMSE and MaxAE for DeepModNet-based and BaseCondNet-based DT models in GSNR estimation.}
\label{fig6}
\end{figure}

\subsubsection{GSNR estimation under domain-aware prediction mode across 35 scenarios}
The GSNR estimation performance under the domain-aware prediction mode is evaluated in this subsection. We substitute the predicted power values from DeepModNet-NLI, DeepModNet-ASE, DeepModNet-Sig, BaseCondNet-NLI, BaseCondNet-ASE and BaseCondNet-Sig into Eq.~(\ref{eq7}) to calculate the estimated GSNR. These estimated GSNR values based on DeepModNet and BaseCondNet are then compared with the ground-truth GSNR to evaluate the estimation accuracy of both approaches across 35 scenarios listed in Table~\ref{tab1}.

\begin{figure*}[t]
\centering
\includegraphics[width=\linewidth]{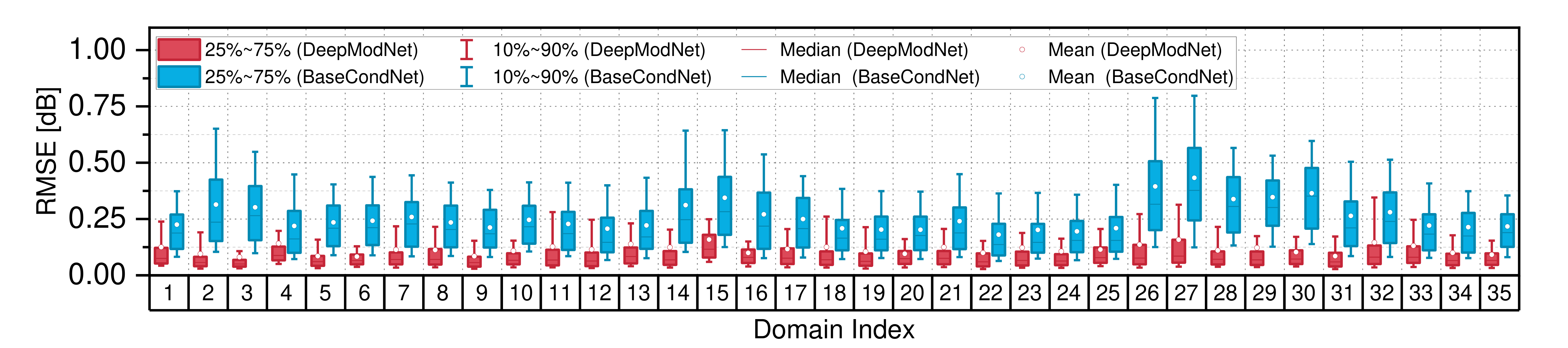}
\caption{Domain-wise box plot distributions of RMSE for DeepModNet-based and BaseCondNet-based DT models in GSNR estimation.}
\label{fig7}
\end{figure*}

Fig.~\ref{fig6} presents the GSNR estimation error distributions over 7,000 samples across the 35 scenarios, showing the PDF of RMSE and MaxAE for both DeepModNet-based and BaseCondNet-based DT models. For DeepModNet, the mean and standard deviation ($\mu \pm \sigma$) of RMSE are $0.114 \pm 0.166~\mathrm{dB}$, while those of MaxAE are $0.197 \pm 0.227~\mathrm{dB}$. In comparison, BaseCondNet-based models produce significantly larger errors, with RMSE values of $0.258 \pm 0.196~\mathrm{dB}$ and MaxAE values of $0.408 \pm 0.263~\mathrm{dB}$. It is evident that DeepModNet significantly outperforms BaseCondNet in GSNR estimation. Using RMSE as the primary metric for estimation accuracy, DeepModNet-based DT models achieve a 55.8\% improvement compared to BaseCondNet. This improvement is primarily attributed to the accurate modeling of signal power by DeepModNet-Sig, which plays a dominant role in GSNR estimation as the signal power is generally much greater than the other two powers during transmission, and the improvement in signal power prediction directly contributes to enhanced GSNR estimation performance.

To further analyze the domain-wise GSNR estimation performance, we present in Fig.~\ref{fig7} the RMSE distributions of DeepModNet-based and BaseCondNet-based DT models across all 35 test scenarios, with each domain containing 200 test samples. Each box plot summarizes the RMSE distribution within a domain: the box spans the interquartile range (25\%$\sim$75\%), the whiskers extend to the 10\% and 90\% percentiles, the horizontal line indicates the median, and the dot denotes the mean.

As shown in Fig.~\ref{fig7}, DeepModNet-based DT models demonstrate superior GSNR estimation accuracy distributions in all of the 35 domains. This trend is highly consistent with the results observed in the earlier power prediction analysis, further demonstrating the effectiveness of the proposed LA-DT in supporting high-precision GSNR estimation. Compared with the baseline, DeepModNet-based models exhibit narrower interquartile ranges, along with lower median and mean RMSE values, indicating smaller estimation errors and more stable performance on the majority of cases. Notably, BaseCondNet-based models display pronounced long-tail distributions in several domains, where the 90th-percentile RMSE values are significantly higher, suggesting larger fluctuations in estimation accuracy for a subset of samples. In summary, DeepModNet-driven LA-DT achieves consistent and accurate GSNR estimation across all domains, demonstrating its robustness and generalization ability across various link conditions.

\subsubsection{GSNR estimation under few-shot fine-tuning mode across 12 unseen scenarios}
The GSNR estimation performance of LA-DT under the few-shot fine-tuning mode for previously unseen scenarios is evaluated in this subsection. Specifically, we compute GSNR values using the predicted NLI, ASE and signal power from corresponding DeepModNet-based DT models before and after fine-tuning. The estimated GSNR values are compared with the ground-truth values for all 2,400 test samples across the 12 domains listed in Table~\ref{tab3}. The evaluation is based on RMSE and MaxAE metrics.

\begin{figure}[t]
\centering
\includegraphics[width= 0.5\linewidth]{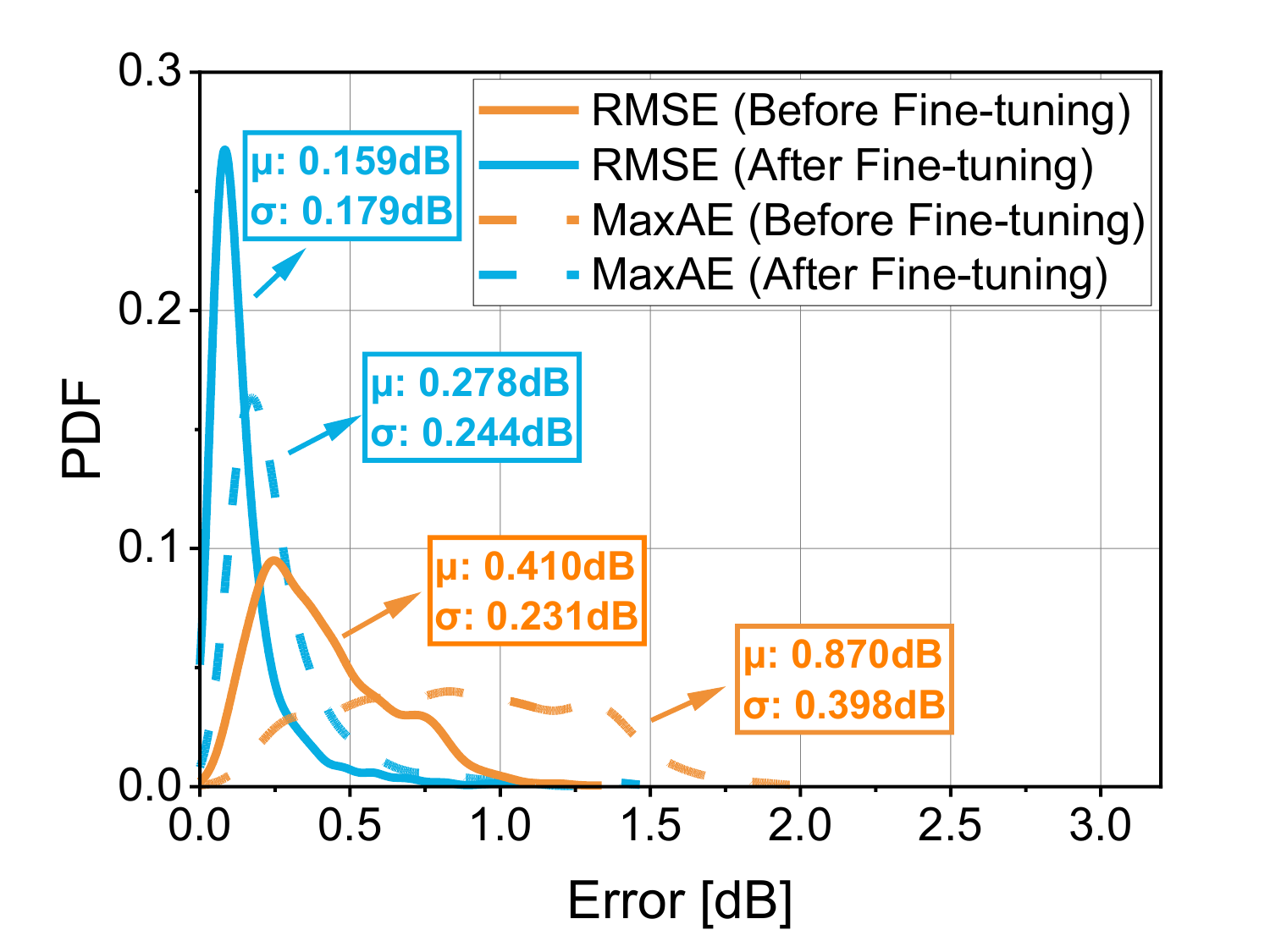}
\caption{Probability density function (PDF) distributions of RMSE and MaxAE for DeepModNet-based DT models before and after fine-tuning in GSNR estimation.}
\label{fig8}
\end{figure}

Fig.~\ref{fig8} shows the PDF distributions of RMSE and MaxAE for DeepModNet-based DT models before and after fine-tuning in GSNR estimation. Before fine-tuning,  the mean and standard deviation ($\mu \pm \sigma$) of RMSE and MaxAE are $0.410 \pm 0.231\mathrm{dB}$ and $0.870 \pm 0.398~\mathrm{dB}$, respectively. After fine-tuning with 20 samples per domain, these values are reduced to $0.159 \pm 0.179~\mathrm{dB}$ for RMSE and $0.278 \pm 0.244~\mathrm{dB}$ for MaxAE. These results demonstrate that even with only 20 fine-tuning samples, the accuracy of GSNR estimation is significantly improved: RMSE is reduced by 61.2\%, and MaxAE decreases by 68.0\%. Such improvements highlight the strong adaptability of the proposed LA-DT in previously unseen domains with limited data. Since GSNR estimation is highly sensitive to the accuracy of power predictions, especially signal power, the enhanced modeling precision achieved through few-shot fine-tuning enables more accurate GSNR estimation.

\begin{figure}[!t]
\centering
\includegraphics[width= \linewidth]{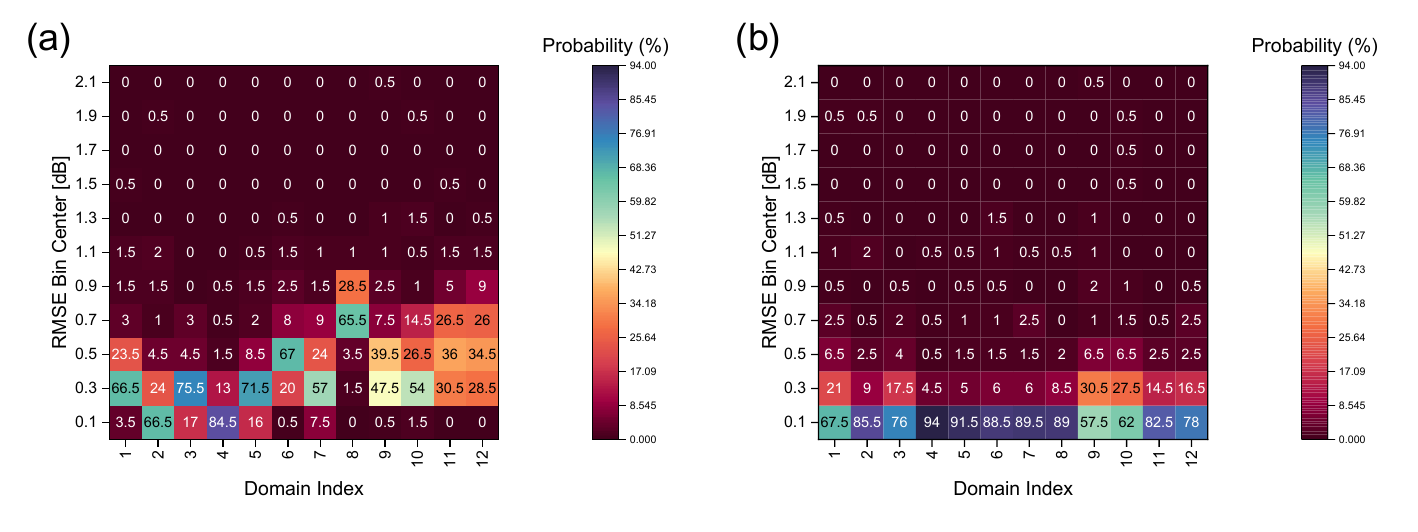}
\caption{Domain-wise heatmap distributions of RMSE for DeepModNet-based DT models before and after fine-tuning in GSNR estimation.}
\label{fig9}
\end{figure}
To further analyze the GSNR estimation performance of LA-DT before and after fine-tuning from a domain-wise perspective, Fig.~\ref{fig9} presents the RMSE distribution heatmaps across 12 unseen domains. In each heatmap, the horizontal axis indicates the domain index, the vertical axis corresponds to the center values of RMSE intervals, and the color intensity represents the probability percentage of each error interval. As shown in Fig.\ref{fig9}, all 12 domains exhibit significant improvements in RMSE distribution after fine-tuning with only 20 samples per domain. After fine-tuning, the RMSE values shift toward lower ranges, with high-probability regions becoming more concentrated in the low-error intervals. In particular, $Domain 8$ exhibits the most noticeable improvement, where the dominant RMSE regions are reduced from 0.6 -- 0.8 dB to less than 0.2 dB. The results demonstrate that the LA-DT is capable of achieving rapid adaptation and substantial accuracy improvements in GSNR estimation for previously unseen domains, even with only 20 samples per scenario, indicating strong adaptability under limited-data conditions.

\section{Conclusion}
\label{Sec6}
In this paper, we proposed the LA-DT framework that enables robust physical-layer modeling in hybrid-amplified ultra-wideband links, achieving both fast inference and strong generalization under diverse link conditions. In LA-DT, GSNR modeling was decomposed into three power prediction tasks to adapt to the heterogeneity of EDFAs in practical ultra-wideband optical networks. Then, three DeepModNet-based DT models were developed equipped with LMLs, enabling domain-aware feature modulation and achieving consistent power modeling accuracy across diverse link conditions involving variations in Raman pump power, launch power, fiber length, and insertion losses. Compared with the baseline general models, these DT models achieved improvements of 56.0\%, 58.4\%, and 52.7\% in NLI, ASE, and signal power predictions, respectively, along with a 55.8\% improvement in GSNR estimation, with an average RMSE as low as 0.114 dB across 35 scenarios. Furthermore, three domain discriminators were designed to guide few-shot fine-tuning of the DeepModNet-based DT models in previously unseen scenarios. Results show that, with only 20 additional samples, this fine-tuning process improves the modeling accuracy of the three powers by 67.7\%, 69.0\%, and 64.8\%, respectively, achieves a 61.2\% improvement in GSNR estimation, and reduces the average RMSE of GSNR estimation to 0.159 dB. In addition, LA-DT explicitly accounts for the variation of insertion loss induced by Raman pumps, an often-overlooked factor that has a significant impact on network performance, thereby ensuring more accurate modeling in practical hybrid-amplified optical networks.

In summary, the proposed LA-DT provides a promising solution for generalized and scalable physical-layer modeling of hybrid-amplified ultra-wideband optical networks. By providing real-time and accurate modeling results, LA-DT can directly support GSNR optimization in such networks and enhance overall network reliability, while also supplying feedback for adaptive resource allocation. \rev{Considering the prevalence of heterogeneous multi-span links in practical networks, future work will focus on extending the proposed framework to support heterogeneous multi-span scenarios by incorporating span-level parameter modeling and mechanisms to capture cascaded power evolution across multiple spans. In addition, incorporating broader system-level factors such as varying channel configurations, modulation formats, traffic loading conditions, and ROADM filtering effects constitutes an important direction for further development of the framework. Integrating these aspects would enhance the practical relevance and applicability of the framework, allowing it to better capture real-world network diversity and complex deployment scenarios.} \rev{Moreover, future work will employ high-precision Split-Step Fourier Method (SSFM)-based simulations and experimental measurements to strengthen the physical rigor of the framework and verify its consistency with fundamental propagation physics under realistic conditions.}

\begin{backmatter}
\bmsection{Funding}
National Natural Science Foundation of China (U25B2013).
\bmsection{Disclosures}
The authors declare no conflicts of interest.
\end{backmatter}

\bibliography{references}
\end{document}